\documentclass[prl,aps,reprint,superscriptaddress,showpacs]{revtex4-1}
\usepackage{graphicx}

\begin{document}

\title{Localised Wannier orbital basis for the Mott insulators GaV$_4$S$_8$ and GaTa$_4$Se$_8$ }

\author{A. Camjayi}
\affiliation{Departamento de F\'{\i}sica, FCEyN, Universidad de Buenos Aires, Ciudad Universitaria, Pab.~1, 1428 Buenos Aires, Argentina.}
\affiliation{IFIBA, CONICET, Ciudad Universitaria, Pab.~1, 1428 Buenos Aires, Argentina.}
\author{R. Weht}
\affiliation{Gerencia de Investigaci\'on y Aplicaciones, Comisi\'on Nacional de Energ\'{\i}a  
           At\'omica (CNEA). Av.~General Paz y Constituyentes, 1650 San Mart\'{\i}n, Argentina.}
\affiliation{Instituto S\'abato, Universidad Nacional de San Mart\'in-CNEA, 1650 San Mart\'in, Argentina.}
\author{M. J. Rozenberg}
\affiliation{Departamento de F\'{\i}sica, FCEyN, Universidad de Buenos Aires, Ciudad Universitaria, Pab.~1, 1428 Buenos Aires, Argentina.}
\affiliation{Laboratoire de Physique des Solides, CNRS-UMR8502, Universit\'e de Paris-Sud, Orsay 91405, France.}

\begin{abstract}
We study the electronic properties of GaV$_4$S$_8$ (GVS) and GaTa$_4$Se$_8$ (GTS), two distant 
members within the large family of chalcogenides AM$_4$X$_8$, 
with A=\{Ga, Ge\}, M=\{V, Nb, Ta, Mo\} and X=\{S, Se\}. 
While all these compounds are Mott insulators,
their ground state show many types of magnetic order, with GVS being ferromagnetic and GTS
non-magnetic.
Based on their bandstructures, calculated with Density Functional Theory 
methods, we compute an effective tight binding Hamiltonian
in a localised Wannier basis set, for each one of the two compounds.
The localised orbitals provide a very accurate representation of the bandstructure,
with hopping amplitudes that rapidly decrease with distance. 
We estimate the super-exchange interactions and show that the Coulomb repulsion
with the Hund's coupling may account the for the different ground states observed
in GVS and GTS.
Our localised Wannier basis
provides a starting point for realistic Dynamical Mean Field Theory studies of
strong correlation effects in this family compounds.
\end{abstract}


\maketitle

The members of the family of chalcogenides compounds of formula AM$_4$X$_8$ (AMX), 
with A=\{Ga, Ge\}, M=\{V, Nb, Ta, Mo\}, X=\{S, Se\}, are paradigmatic 
examples of strongly correlated systems.
Their recent experimental study has unveiled very interesting properties such as pressure
driven metal-insulator transitions, superconductivity above 11.5 GPa ~\cite{prl2004}, 
and even resistive switching under electric pulsing~\cite{vaju2008,inoue-rozenberg}.

Crystallographically, these compounds have a lacunar spinel structure with a FCC 
general symmetry~\cite{barz, perrin}.
They are formed by two types of units: clusters of the transition metal atoms surrounded 
by the X atoms, M$_4$X$_8$, and AX$_4$ tetrahedrons, both ordered in a NaCl manner.
As the distances between the four metal atoms of the cluster are significantly 
shorter than the inter-cluster ones we can understand
their basic electronic properties on the basis of molecular orbitals. 
Particularly, near the Fermi energy they form three molecular levels 
with a $t_{2g}$ cubic symmetry, which may be filled with one or two electrons, 
or even with one hole, depending on the specific combination of the elements above ~\cite{pocha2000}.
The smaller overlap between inter-cluster orbitals gives place to 
relatively small hopping amplitudes which lead to narrow $d$-electron bands, 
where strong correlation effects may be expected.
In fact, the nominal occupation of the molecular orbitals should lead to partially 
filled metallic bands, however, all the experimental systems are good insulators at low temperatures. 

There is another universal feature that runs through the whole family, namely, the 
occurrence of structural transitions at temperatures around 50~K. 
However, not all compounds adopt the same low temperature structure, and several of 
those low-$T$ structures remain yet undetermined.

Regarding their ground-state properties, there are also striking differences. 
While some of them acquire a magnetic order, such as GaV$_4$S$_8$ and 
GeV$_4$S$_8$, which respectively become a ferro- and an antiferro-~magnet~\cite{pocha2000, johrendt}, 
others, such as GaNb$_4$S$_8$ and GaTa$_4$Se$_8$, seem to remain paramagnetic 
down to the lowest measured temperatures.

Since all these systems have uncompleted electronic shells but fail to become metals,
they are generally considered to be a family of paradigmatic Mott insulators. 
Moreover, experimentally, it has been observed that pressure may dramatically decrease
the resistivity of all these compounds, up to several orders of magnitude at low temperature, 
consistent with this classification.
Some of them may even become superconductors under pressure, as for instance 
GaTa$_4$Se$_8$, with a critical temperature of just a few degrees K at 11.5 GPa~\cite{prl2004}.
The Mott transition in the AMX family has also been associated to their
unusual resistive switching properties,
which may be used for novel non-volatile memory devices~\cite{vaju2008,inoue-rozenberg}.

While the universal classification of these insulators as of Mott-type is appealing,
it also poses the question on the origin of their observed differences. 
In this paper we begin to address this issue by obtaining a representation of the electronic 
structure in terms of localised orbitals, and studying the systematic differences 
by focusing in two distant members of the family, namely, GaV$_4$S$_8$ (GVS) and GaTa$_4$Se$_8$ (GTS). 
The former has more localised 3$d$ electrons while the latter has more extended 5$d$ ones. 
The possibility of representing the electronic bandstructure through a tight binding Hamiltonian
is important in two ways: Firstly, it justifies that by adding the correlation effects one would obtain
a Hubbard-type model where Mott physics takes place in the strongly correlated limit, thus
accounting for the universal insulator behaviour in the intermediate to high (room) temperature range.
Secondly, the inter-cluster hopping matrix elements obtained in the tight binding construction 
are closely connected with the effective short-ranged magnetic interactions, such as the 
super-exchange mechanism. Therefore, their systematic changes may provide insights on the 
origins of the variety of ground-states that are experimentally observed.

Our results show that the electronic structures of both studied compounds
can in fact be faithfully reproduced by an effective tight binding Hamiltonian defined on a
localised orbital basis set, which we compute explicitly.  The computed data for
the two systems seem qualitatively similar, which is consistent with the observed universal Mott behaviour
at intermediate to high temperature. However, the inter-cluster hopping amplitudes and the estimated 
super-exchange interactions reveal significant differences, which
may explain the variety of ground-states that are also experimentally observed. 
In addition, our work also provides a suitable starting point for a Dynamical Mean Field theory 
(DMFT) study~\cite{review}, which may eventually fully elucidate the detailed behaviour of the members of the AMX
family.

Most of the previous theoretical work on compounds of this family has been restricted to 
Density Functional Theory approaches~\cite{pocha2005, muller2006},
in general including correlation effects that are introduced as a static mean field, 
such as in the LDA+U methods~\cite{anisimov}.
However, unless a low temperature structural transition is assumed, and a certain 
type of magnetic ordering is adopted, these approaches would usually fail to predict 
insulator states.
Other approaches, such as the DMFT should be better adapted, as they 
may predict paramagnetic or magnetic Mott insulator states on equal footing. 
However, that methodology would also rely on two assumptions, namely the value of the
onsite Coulomb interaction strength $U$, and the definition of a suitable localised basis
set. Here we shall explicitly obtain such a localised basis for the two compounds previously
mention, and we shall discuss their main similarities and differences.

The starting point of our approach is the Density Functional Theory (DFT)~\cite{dft} 
electronic structure calculation of the compounds GVS and GTS.
To this end we use the Wien2K code~\cite{wien2k}, that is an implementation of
the full-potential linearised-augmented plane wave method (FP-LAPW)~\cite{lapw}.
As we are not interested in total energies nor magnetic states, we adopt the simplest 
local density approximation to represent the exchange-correlation potential~\cite{lda}.
Results using the generalised gradient approximation are completely equivalent
since band structures are, in general, insensitive to this choice.
Even thought within the DFT schemes the eigenvalues formally have no direct physical meaning, 
it is, nevertheless, broadly accepted that they provide a good approximation for the quasi-particle 
energies and band-structure.  Thus, here we shall adopt them as the reference for the calculation of the localised
Wannier orbital basis. We restrict the energy window to the three $t_{2g}$ bands that
cross the Fermi energy. We note that these three bands are quite well separated from the rest of the
band manifold, which is very advantageous in order to successfully obtain an accurate 
local orbital basis representation with short ranged hopping amplitudes.
The localised Wannier orbital basis is computed following the procedure described by
Marzari, Vanderbilt and coworkers in a series of papers~\cite{MLWF} and implemented in
the code Wannier90~\cite{W90}. As the interface between the two programs we use 
the code wien2wannier~\cite{w2w}.

\begin{figure}
\includegraphics[width=\linewidth]{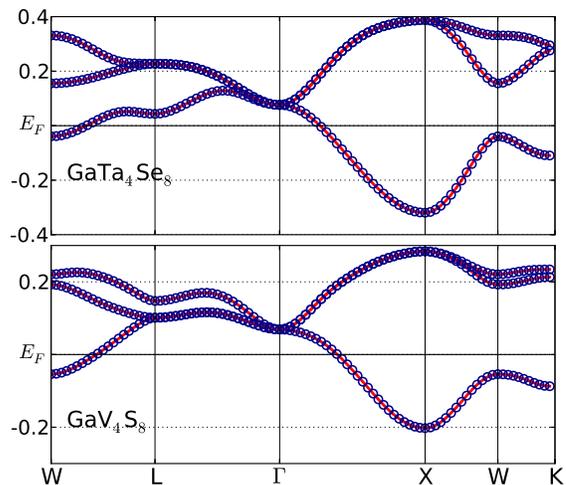}
\caption{(Color online) The red curves are the DFT electronic structure 
of the $t_{2g}$ bands crossing the Fermi energy for GVS (bottom) and
GTS (top). The blue-dot curves correspond to the respective tight 
binding energy bands.}
\label{fig.1}
\end{figure}

In fig.~\ref{fig.1} we show the electronic band-structure obtained by the DFT calculations
along with those computed from the effective tight binding Hamiltonian on the localised
Wannier orbital basis. We observe that the agreement is excellent for both, the  
GVS and the GTS compounds.

\begin{figure}
\includegraphics[width=0.7\linewidth]{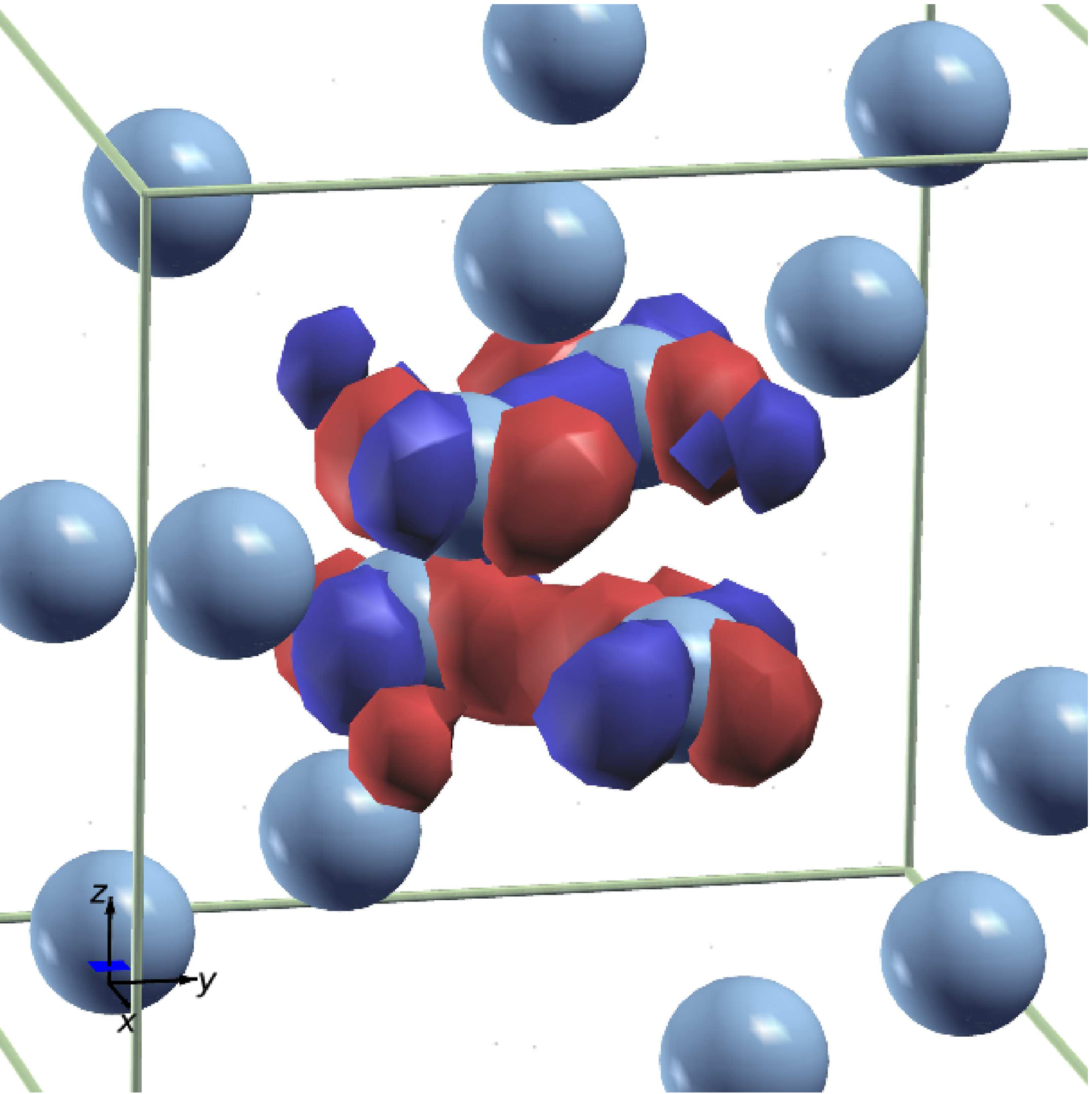}
\includegraphics[width=0.9\linewidth]{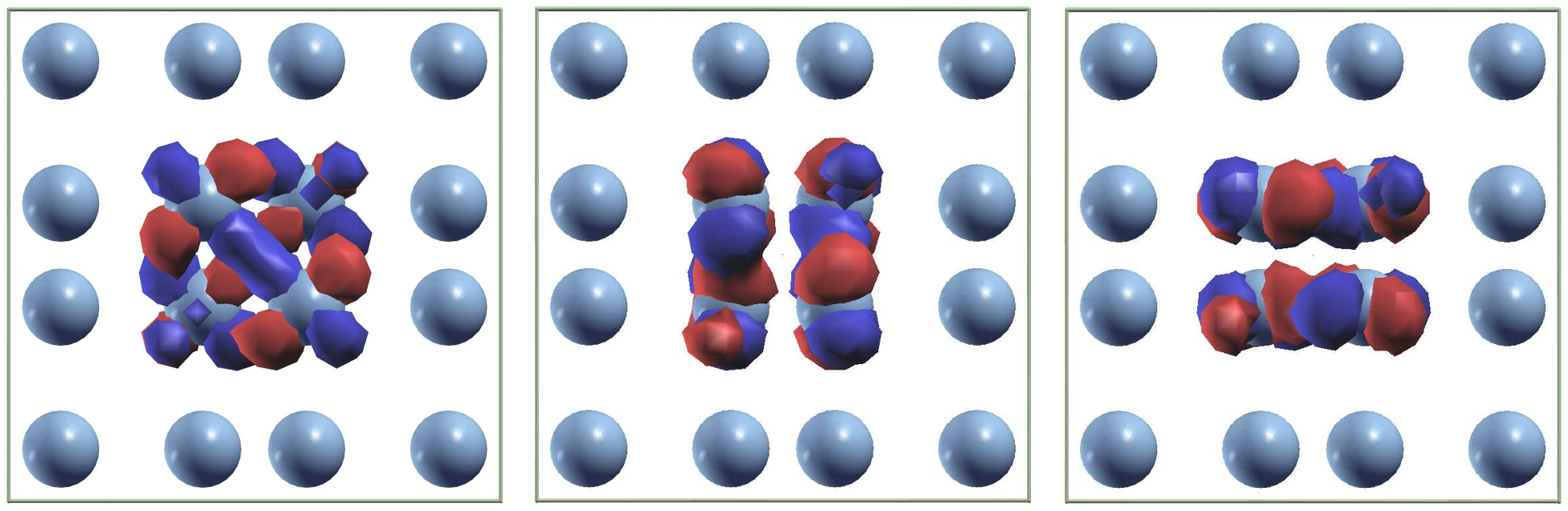}
\caption{(Color online) Top panel: Spatial representation of one of the three
$t_{2g}$ Wannier orbital of GVS. Bottom panels:
Side view (100) of the three Wannier orbitals $d_{xy}$,  $d_{yz}$ and $d_{zx}$ of GVS.
For the sake of clarity, Ga and S atoms are not shown.
The results for the corresponding orbitals of the GTS compound are 
qualitatively similar, hence, they are not displayed.}
\label{fig.2}
\end{figure}

In fig.~\ref{fig.2}, we depict the Wannier orbitals constructed for GVS.
These orbitals have the $t_{2g}$ symmetry expected for the cubic lattice.
However, they are molecular orbitals involving the four V atoms forming
a cluster, thus, as can be seen in the figure, they are different 
from the familiar single-atom $d$-shell $t_{2g}$ symmetry orbitals.
The comparison of the two sets of Wannier orbitals computed for GVS and GTS show that
the former are relatively more confined. The spread of the orbital, taken as a representative of the 
extension of the wave function, is around 7.5~\AA\ for GVS but slightly more than 10~\AA\ for GTS.

\begin{figure}
\includegraphics[width=\linewidth]{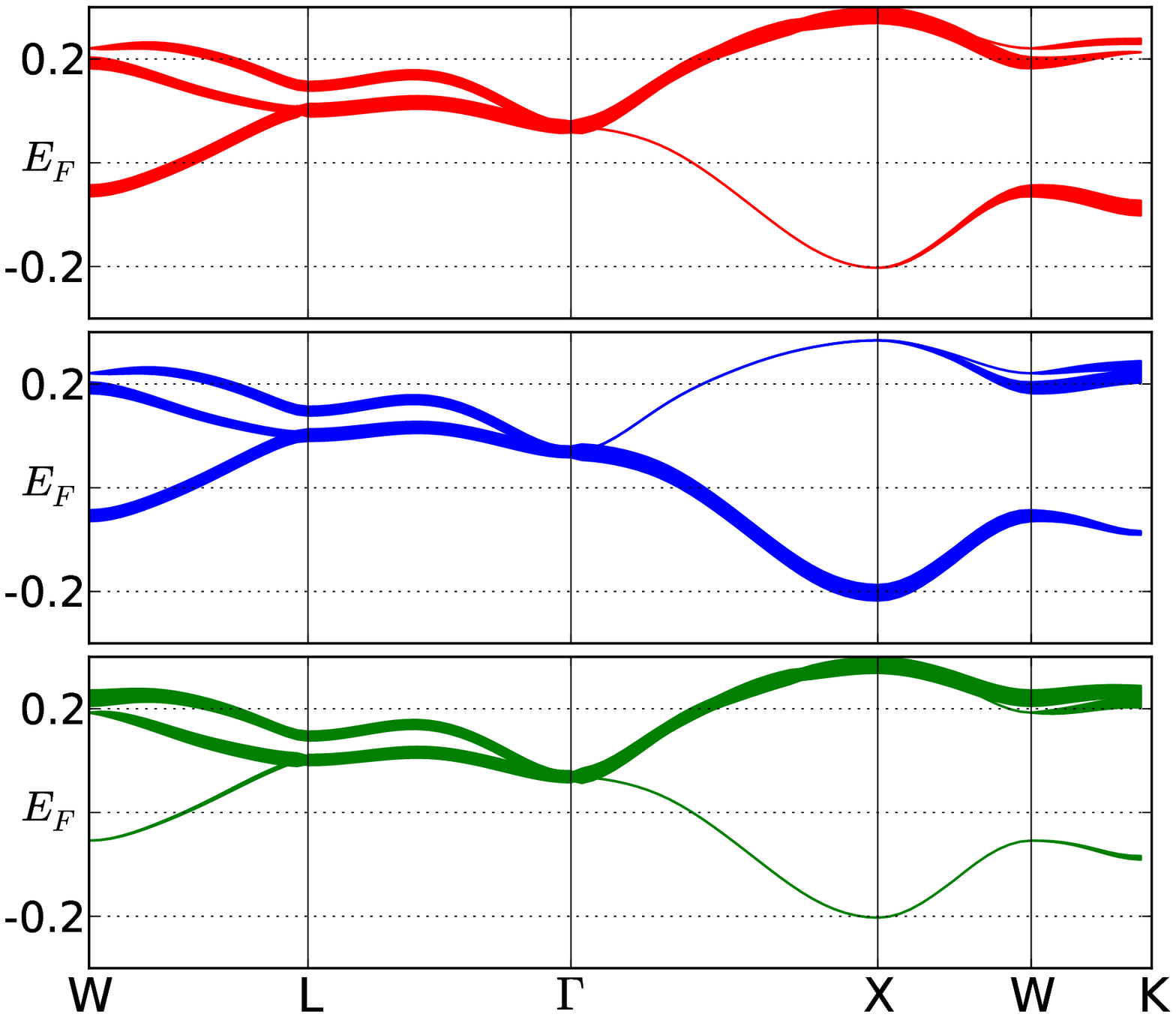}
\includegraphics[width=\linewidth]{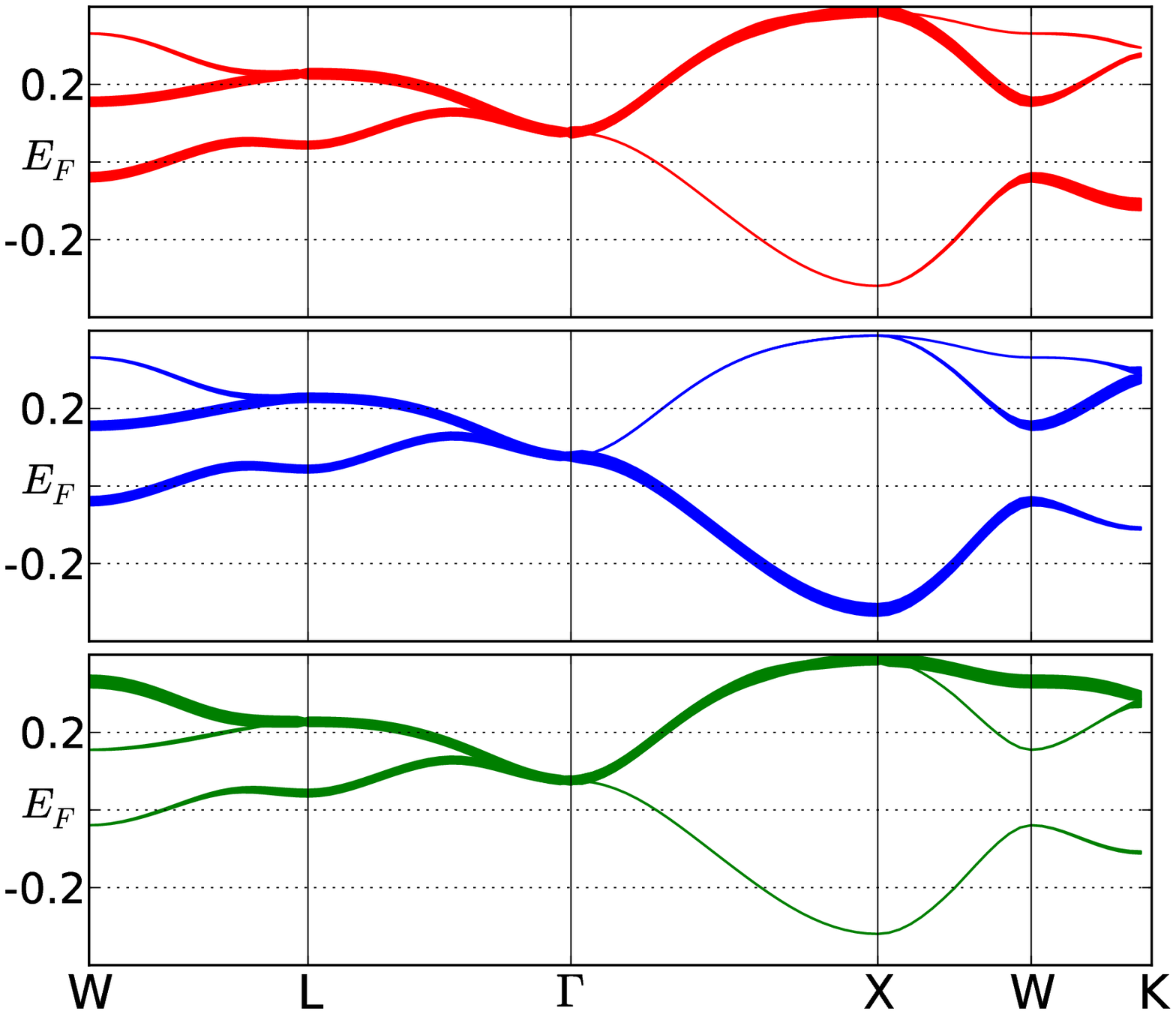}
\caption{(Color online) ``Fat bands'' for GVS (top) and GTS (bottom). Red, green and blue 
respectively correspond to $d_{xy}$,  $d_{yz}$ and $d_{zx}$ characters.}
\label{fig.3}
\end{figure}

In fig.~\ref{fig.3} we show the ``fat bands'', which contain the information of the orbital content
of each one of the bands. We observe that the GVS compound shows a relatively high
degree of orbital character mixing in the three bands, while the GTS compound, in contrast, 
shows a smaller mixing.

\begin{figure}
\includegraphics[width=\linewidth]{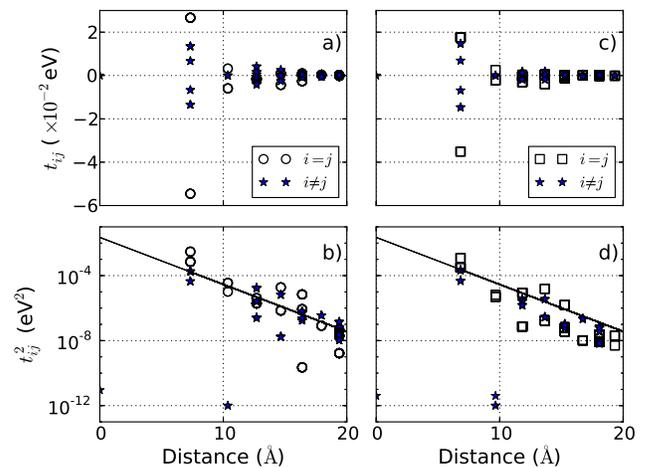}
\caption{(Color online) Orbital hopping amplitudes and square of the same quantity for GTS (a, b) and GVS (c, d). Open symbols are for same-orbital and full symbols for different-orbital hopping.
The lines on the bottom plots are a guide for the eyes.}
\label{fig.4}
\end{figure}

In fig.~\ref{fig.4} we show the dependence of the orbital hopping elements
$t_{ij}$ with distance. Notice that by symmetry 
$t_{xy,xy}({\bf r}) =  t_{yz,yz}({\bf r}) = t_{zx,zx}({\bf r})$, and
$t_{xy,yz}({\bf r}) =  t_{yz,zx}({\bf r}) =  t_{zx,xy}({\bf r})$  .
The results show that for both GVS and GTS the magnitude of the overlaps decrease very
rapidly with distance. They become virtually negligible beyond the first nearest neighbours for GVS,
while in GTS they reach out a little longer.
Moreover, quantitatively, the hopping elements of GVS are significantly smaller than those of GTS, 
which is consistent with the smaller bandwidth of the former.

The calculation of the $t_{ij}$, shown in fig.~\ref{fig.4}, 
enable us to gain further insight on the possible
origin of different ground-states observed in GVS and GTS. If we consider the fact that
both systems are in the large-$U$ Mott insulator regime, their conduction band electrons 
become actually localised by correlations, one occupying each transition-metal tetrahedron. 
These localised states are highly degenerate since the electron
may occupy any one of the three $t_{2g}$ molecular orbitals with any of the two spin directions.
In the simpler case of a one band Mott-Hubbard insulator in a bipartite lattice, 
this type of degeneracy would be lifted by the superexchange interaction ($\sim t^2/U$), 
which drives the system towards an antiferromagnetically ordered ground-state.
Here the situation is more complex, due to the higher degeneracy and the more complex
lattice structure. Nevertheless, in a first approximation, the examination of the 
generalised superexchange interaction terms should give clues on the different nature of
the ground-states in the different materials.

The generalisation of the superexchange interactions 
to the present multiorbital case is straight forward ~\cite{pavarini}.
The magnitude of the couplings are $J^{se}_{i,j} = t^2_{ij}/(U-\nu J)$, where $J$ is
Hund's interaction and $\nu$= 3, 2 or 0, depending whether the doubly occupied
virtual state has different orbital and same spin, same orbital and different
spin, or different orbital and different spin, respectively~\cite{pavarini}.
The first case, $\nu = 3$, is the one that may favour a localised
ferromagnetic insulator ground-state. We note that the stabilisation of a ferromagnetic
state would imply the (anti-)orbital ordering of the electronic states in the lattice.
From the relative values of the $t_{ij}$ that are plotted in fig.~\ref{fig.4}, 
it is immediately apparent that the 
superexchange interactions are dominated by the nearest-neighbour same-orbital
terms (\textit{i.e.}, $i=j$).
An important uncertainty that we face are the values of the interactions $U$ and $J$.
A first principle methodology for their calculation remains a mater of debate.
Here we shall take a practical approach.
The value of the interaction $U$ may be estimated from optical conductivity
experiments, which probe direct lower to upper Hubbard band transitions. The corresponding
values for GVS and GTS are 0.8 and 1.2 eV \cite{vinh}. These are in fact larger than the
respective bandwidth $W$, 0.5 and 0.7 eV, and the ratio $U/W$ is approximately 1.6 and 1.71
for GVS and GTS, respectively. These values should be taken as lower
bounds for the estimate of the strength of the interaction.
%If we takeonly the dominant band crossing the Fermi energy, we get even higher ratios.

On the other hand, from studies of DMFT in multi-orbital systems\cite{ono}, 
it is known that the Mott insulator at $T=0$
occurs for a critical $U_c$ of approximately $10t$, for a three degenerate-band Hubbard
model with one electron per site. In the case
of a single band model with one electron per site, the value of $U_c$ 
about $6t$. The total bandwidth for the three degenerate bands is $W = 4\sqrt{3}t$,
while for the single band case is $W = 4t$.
Hence, $U_c/W \approx 1.44$ for the three-band case, and
$U_c/W \approx 1.5$ for the single band case.
The critical value for GTS and GVS should be somewhere in between these
two cases, since the three-band degeneracy is partially lifted.
Clearly, from the considerations made above, from DMFT one would expect
both GVS and GTS to be Mott-Hubbard insulators.

\begin{figure}
\includegraphics[width=\linewidth]{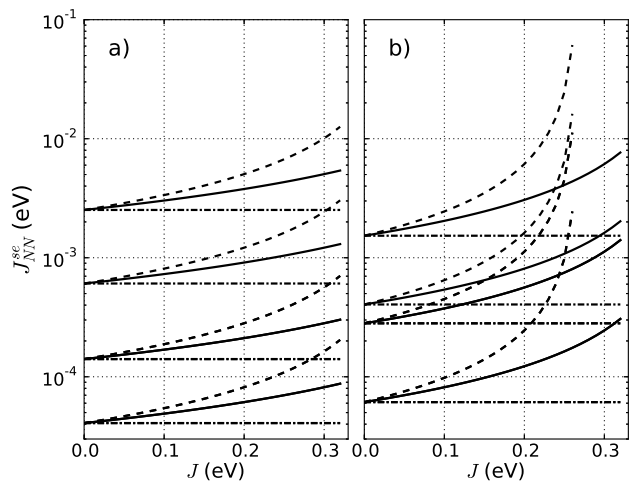}
\caption{First near-neighbour superexchange interaction for GTS (a) and GVS (b). 
The cases with $\nu = 0, 2$ and 3 (see text) are, respectively, dashed-dotted, 
full and dashed lines. In both cases, the two topmost sets correspond to 
same-orbital exchange ($i=j$) and the rest to the different-orbital case ($i\neq j$).}
\label{fig.5}
\end{figure}

The estimation of $J$ is more difficult, so we shall simply
consider it a free parameter. Reasonable values of $J$, however, would run up to
about a third of the interaction $U$.
In fig.~\ref{fig.5} we show the dependence of the superexchange
couplings for nearest neighbour sites at fixed interaction $U$ and as a function of the Hund's parameter $J$.
From the figure results it is apparent that the GVS system has a stronger
tendency towards ferromagnetic order, since the $\nu = 3$ superexchange interaction
is larger than the others, and relatively larger than in GTS. This result
is consistent with the experimentally determined groudstates.

While these observations are suggestive, we should emphasise, nevertheless, 
that the present study, based on a strong coupling picture, remains inconclusive.
In fact, an interesting open question, for instance, is what would be the actual orbitally 
ordered state that would correspond to the ferromagnetic insulating Mott state of GVS.
Also, to go beyond this strong coupling approach, one would need to perform a fully 
quantum mechanical many-body calculation, such as in realistic Dynamical Mean Field Theory.
Our calculation and our effective tight binding Hamiltonian should be a proper
start point for such a calculation, which we plan for future work.
An additional complication that we have so far ignored is the fact that GVS, unlike
GTS has a structural transition to a R3m structure that occurs before the ferromagnetic 
instability\cite{yadav}. It remains an open question but, is in fact possible, 
that the ferromagnetic instability is further
favoured by the distortion.

To conclude, we have studied the electronic states of GVS and GTS,
which belong to a large family of correlated chalcogenides that 
are commonly classified as Mott insulators. These systems share the
same high temperature structure, however,  they
display different magnetic ground-states. By means of density functional
calculations and a maximally localised Wannier orbital construction,
we explored the physical origin of the observed differences.
We considered GVS and GTS as two test-case systems, since the first
orders ferromagnetically while the second does not seem to order.
From the orbital construction we computed the hopping amplitudes and 
obtained the effective superexchange
interactions within a strong coupling picture.
We found quantitative differences,
consistent with the experimental observations.

\acknowledgments
RW gratefully acknowledge support from CONICET and ANPCyT (grant PICT 837/07).

\end{document}